\documentclass[twocolumn,english,aps,superscriptaddress,showpacs]{revtex4-1}
\usepackage[T1]{fontenc}
\usepackage[latin9]{inputenc}
\usepackage{bm}
\usepackage{amsmath}
\usepackage{amssymb}
\usepackage{graphicx}
\usepackage{esint}

\makeatletter

\providecommand{\tabularnewline}{\\}

\usepackage{mathrsfs}
\usepackage{amsfonts}
\usepackage{subfigure}
\usepackage{bm}
\usepackage{verbatim}
\usepackage{color}
\usepackage{xcolor}
\usepackage{epsfig}
\usepackage{longtable}

\newcommand{\upperRomannumeral}[1]{\uppercase\expandafter{\romannumeral#1}}

\makeatother

\usepackage{babel}
\begin{document}

\title{Evolution of the optimal trial wave function with interactions in
fractional Chern insulators }

\author{Yumin Luan}

\affiliation{International Center for Quantum Materials, Peking University, Beijing
100871, China}

\author{Yinhan Zhang}

\affiliation{International Center for Quantum Materials, Peking University, Beijing
100871, China}

\author{Junren Shi}
\email{junrenshi@pku.edu.cn}

\affiliation{International Center for Quantum Materials, Peking University, Beijing
100871, China}

\affiliation{Collaborative Innovation Center of Quantum Matter, Beijing 100871,
China}

\date{\today}
\begin{abstract}
We show that the optimal trial wave function of a fractional Chern
insulator depends on the form of its electron-electron interaction.
The gauge of single particle Bloch bases for constructing the optimal
trail wave function is obtained by applying the variational principle
proposed by Zhang \textit{et al.} {[}Phys. Rev. B \textbf{93}, 165129
(2016){]}. We consider a short-range interaction, the Coulomb interaction,
and an interpolation between them, and determine the evolution of
the optimal gauge with the different interactions. We compare the
optimal gauge with those proposed by Qi {[}Phys. Rev. Lett. \textbf{107},
126803 (2011){]} and Wu \textit{et al.} {[}Phys. Rev. B \textbf{86},
085129 (2012){]}, and find that Wu \textit{et al.}'s gauge is close
to the optimal gauge when the interaction is a certain mixture of
the Coulomb interaction and the short-range interaction, while Qi's
gauge is qualitatively different from the optimal gauge in all the
cases. Both the gauges deviate significantly from the optimal gauge
when the short-range component of the interaction becomes more prominent.
\end{abstract}
\maketitle

\section{Introduction}

Fractional quantum Hall effect (FQHE), which exhibits fractional plateau
of the Hall conductance in high magnetic field and low temperature~\cite{Tsui},
is one of the most important discoveries of condensed matter physics.
Different from the single particle nature of the integer quantum Hall
effect (IQHE)~\cite{Klitzing1980,TKNN}, the FQHE is driven by electron-electron
interaction~\cite{Laughlin1983,Jain1989,JainCF,Haldane1983}. Actually,
it is the first topological state ever discovered that is induced
by an interaction. For the reason, the study of the effect occupies
a center position in theoretical inquiries of condensed matter physics.
Moreover, some of FQH states could even find potential applications
in topological quantum computing because they support excitations
of non-Abelian statistics, which could be utilized to encode quantum
information free from local disturbances~\cite{MR,NayakRMP}. 

Recently, theoretical studies reveal a new class of lattice models
that exhibit FQH states without a magnetic field~\cite{Sheng,SunKai,Neupert,Tang,Wangyifei}.
All of these models possess at least a flat Chern band that is nearly
dispersionless and isolated from other bands by energy gaps, and is
topologically nontrivial with a nonzero Chern number. The band imitates
a Landau level in ordinary FQH systems. Similar to the case of a Landau
level, in the presence of an electron-electron interaction, the band
with a fractional filling factor could also exhibit the FQHE, resulting
in a fractional Chern insulator (FCI)~\cite{Bernevigprx,Parameswaran,LiuZhaoreview,NeupertFTI}.
The important and remarkable feature of FCIs is that they can potentially
be realized in zero magnetic field and high temperature~\cite{Neupert,Tang},
which is highly desirable for applications. 

In order to understand the FQH physics arisen in FCIs, it is important
to find a way to construct their many-body ground state wave functions,
as Laughlin's wave function for ordinary FQH systems~\cite{Laughlin1983}.
To this end, Qi proposes a mapping approach which obtains the ground
state wave function of a FCI from a FQH wave function of the same
filling fraction~\cite{Qi2011}. Specifically, one expands a FQH
many-body wave function in single-body Landau orbitals. By replacing
the single-body Landau orbitals with a set of single-body bases constructed
in a FCI, we obtain a ground state trial wave function for the FCI.
Unfortunately, the mapping method suffers from the arbitrariness of
the choices of the Landau orbitals and the FCI bases, as well as the
correspondence between them. In this aspect, Qi chooses LLL orbitals
in the Landau gauge on a cylinder, and map them to a set of one-dimensional
localized Wannier functions constructed in the flat Chern band of
a FCI~\cite{Qi2011}. Wu \textit{et al.} adopt an alternative mapping
by considering the effect of finite-size and analogousness of phase
between LLL orbitals and Wannier orbitals of a FCI. It achieves a
higher overlap with the exact ground state wave function of a FCI
compared to Qi's approach~\cite{Wu2012,Wu2013,Wu2014}.

Zhang \textit{et al.} indicate that the arbitrariness is actually
the choice of a gauge for constructing two-dimensional (2D) localized
Wannier functions when mapping a continuous system to a lattice model.
From the observation, they establish a general variational principle
for determining the optimal gauge that minimizes the interaction energy~\cite{Zhang2016}.
An immediate consequence from the consideration is that the optimal
gauge should depend on the form of the interaction adopted in a FCI
model. This is in sharp contrast with Qi's or Wu \textit{et al}.'s
approaches, both of which prescribe a single mapping for all possible
FCI models derived from one lattice model but with different forms
of interaction. While the general principle is established in Ref.~\cite{Zhang2016},
its manifestation in a real FCI model was not explicitly demonstrated.
It would be desirable to see how the different forms of interaction
affect the optimal gauge, and how Qi's and Wu \textit{et al}.'s choices
of the gauge are compared to the optimal one.

In this paper, we demonstrate the dependence of the optimal gauge
(or equivalently, the optimal trial ground state wave function) on
different forms of electron-electron  interaction in the checkerboard
model~\cite{Sheng,SunKai,Neupert,Wangyifei}. The optimal gauge is
determined by the interaction energy variational principle proposed
by Zhang \textit{et al.}~\cite{Zhang2016}. We consider three forms
of interaction, including a short-range interaction which is widely
adopted in literatures, the Coulomb interaction, and an interpolation
between them. We find that, when varying the form of interaction,
the optimal gauge changes significantly. The corresponding Wannier
functions, which facilitate the mapping from Landau levels in continuous
space to the lattice model, also change in both spatial distribution
and symmetry. We compare the optimal gauge with those determined by
Qi's proposal and Wu \textit{et al}.'s proposal. We find that Wu \textit{et
al.}'s gauge can be close to the optimal gauge when the interaction
is a certain mixture of the Coulomb interaction and the short-range
interaction, while Qi's gauge is qualitatively different from the
optimal gauge with a different spatial symmetry for all the cases.
Both the gauges deviate the optimal one significantly when the short-range
component of the interaction becomes more prominent.

The remainder of the article is organized as follows. In Sec.~II,
we present the tight-binding model of the checkerboard lattice, and
three forms of the interaction surveyed in the present work are introduced.
The method for determining the optimal gauge as well as its numerical
implementation are discussed. In Sec.~III, we present the results
of the optimal gauges for the three forms of interaction. Finally,
Sec.~IV contains a concluding remark.

\section{Model and methods}

\subsection{Lattice model}

\begin{figure}[tb]
\centering \includegraphics[width=0.4\columnwidth]{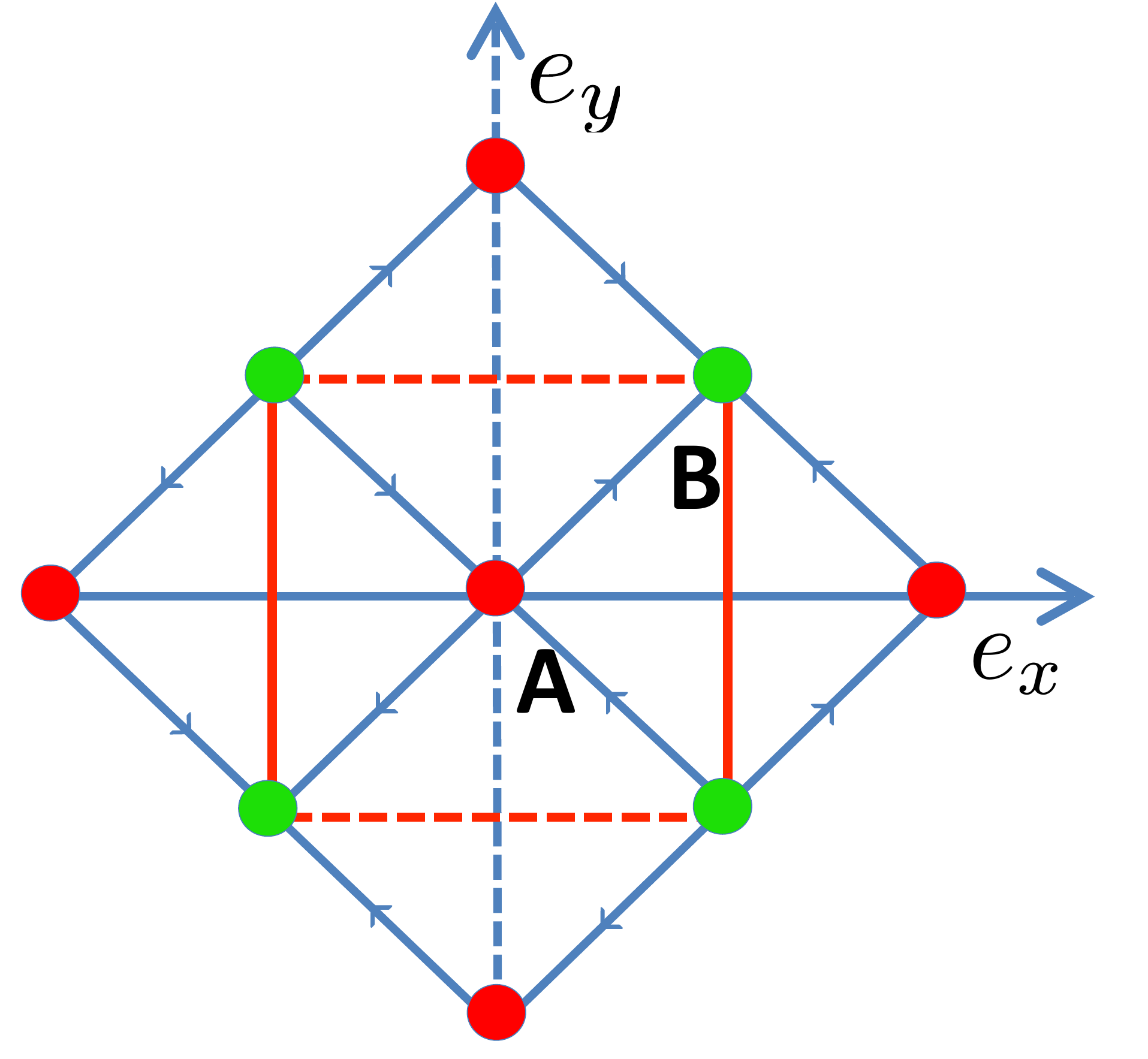} \includegraphics[width=0.57\columnwidth]{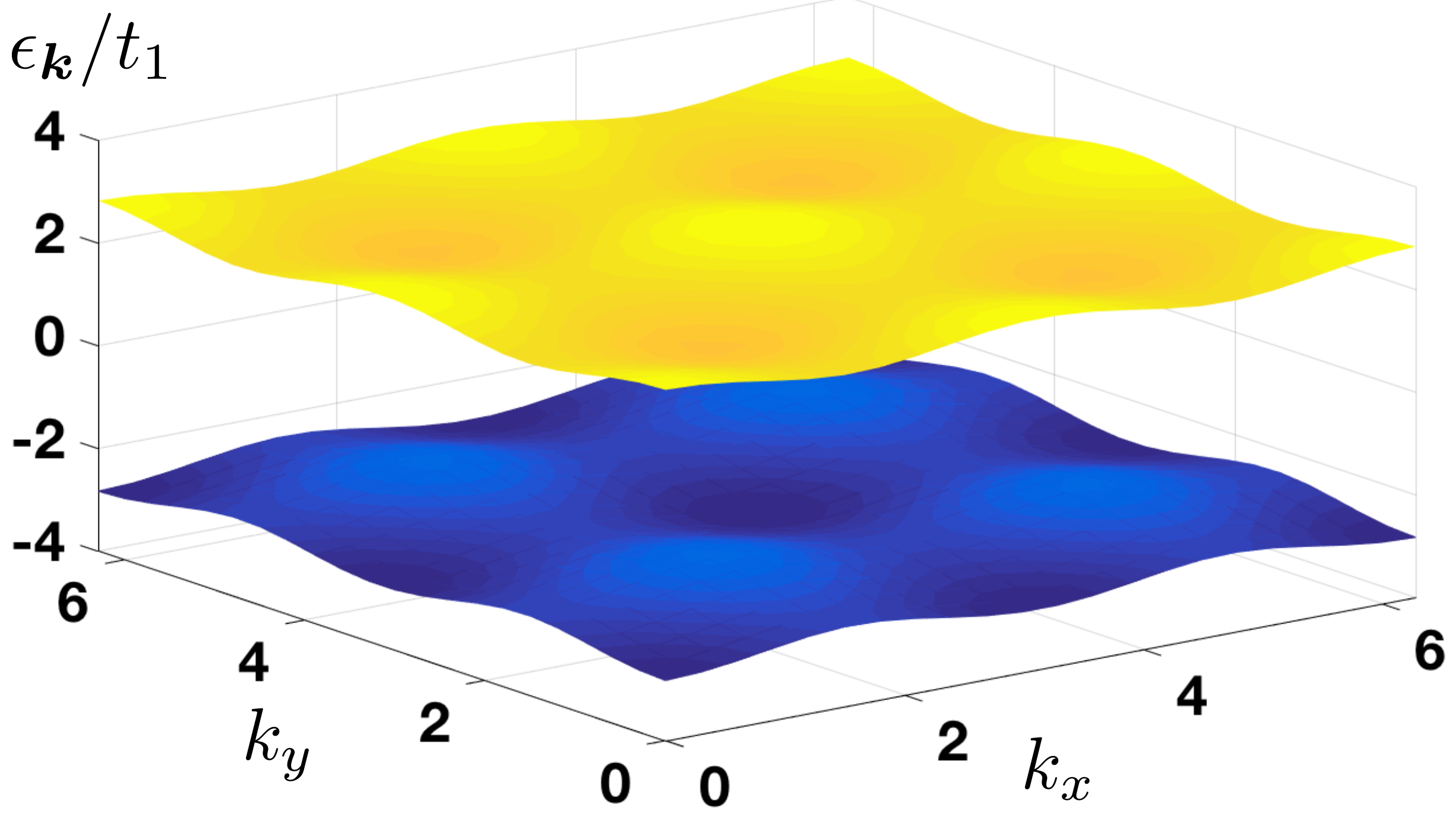}
\caption{(Color online) Left: Lattice configuration of the checkerboard model.
Each unit cell, which is defined as the area enclosed by the red lines,
contains two sites A (red circles) and B (green circles). Lattice
vectors are $\bm{a}_{x}\equiv(1,0)$ and $\bm{a}_{y}\equiv(0,1)$.
The NN hopping amplitudes represented by arrows direction is $t_{1}\textrm{exp}(i\pi/4)$,
and the NNN hopping amplitudes, represented by dashed lines and solid
lines, are $-t_{2}$ and $t_{2}$, respectively. Right: The flat Chern
bands of the checkerboard lattice model with parameters $t_{1}=\sqrt{2}t_{2}$.}

\label{fig: TBModel}
\end{figure}

A large number of flat Chern band models of various lattice configurations
had been proposed in literatures. These include models with a checkerboard
lattice~\cite{Sheng,SunKai,Neupert,Wangyifei}, a kagome lattice~\cite{Tang},
and a honeycomb lattice~\cite{Wangyifei}, all of which possess a
flat band with a Chern number $C=\pm1$. Models with a higher Chern
number had also been proposed in such as pyrochlore slabs~\cite{LiuzhaoHighC,Sterdyniak,Trescher},
dice lattice~\cite{Wangfa}, and triangle lattice~\cite{WangyifeiHighC,Yangshuo,Cooper}.
In this paper, for simplicity, we choose the checkerboard lattice
model as our system for demonstrating the interaction dependence of
the optimal trial wave function of FCIs. 

The lattice configuration of the checkerboard model is shown in the
left panel of Fig.~\ref{fig: TBModel}. The tight-binding Hamiltonian
of the model with nearest-neighbor (NN) and next-nearest-neighbor
(NNN) hopping terms can be expressed, in the reciprocal space, as~\cite{Neupert}:
\begin{gather}
H=\sum_{\bm{k}}{\psi_{\bm{k}}^{\dagger}\mathcal{H}_{\bm{k}}\psi_{\bm{k}}},\quad\quad\mathcal{H}_{\bm{k}}=\bm{T}(\bm{k})\cdot\bm{\sigma},\nonumber \\
T_{1}(\bm{k})-iT_{2}(\bm{k})=t_{1}e^{-i\frac{\pi}{4}}[1+e^{i(k_{y}-k_{x})}]\nonumber \\[3mm]
+t_{1}e^{i\frac{\pi}{4}}[e^{-ik_{x}}+e^{ik_{y}}],\\
T_{3}(\bm{k})=2t_{2}(\text{cos}k_{x}-\text{cos}k_{y}),
\end{gather}
where $\psi_{\bm{k}}\equiv(\alpha_{\bm{k}A},\alpha_{\bm{k}B})^{T}$
with $\alpha_{\bm{k}\tau}$ being the annihilate operator for a state
with a wave vector $\bm{k}$ and at the sub-lattice $\tau=A,\,B$,
and $\bm{\sigma}\equiv(\sigma_{1},\sigma_{2},\sigma_{3})$ with $\sigma_{i},\,i=1,2,3$
being the Pauli matrices. $T_{1}(\bm{k})$ and $T_{2}(\bm{k})$ correspond
to the NN term between A site and B site, while $T_{3}(\bm{k})$ is
the NNN hopping term in the same sub-lattice, and $t_{1}$ and $t_{2}$
are respective hopping constants. When $t_{1}=\sqrt{2}t_{2}$, the
bands become flattest~\cite{Neupert}. The right panel of Fig.~\ref{fig: TBModel}
shows the band structure under this condition.

One can diagonalize the Hamiltonian, and obtains two eigenvalues $\epsilon_{1,2}(\bm{k})=\mp|\bm{T}(\bm{k})|$
as well as eigenvectors $u_{1,2}(\bm{k})$: 
\begin{gather}
u_{1}(\bm{k})=\begin{pmatrix}e^{-\frac{i}{2}\gamma_{\bm{k}}}\text{sin}\frac{\varphi_{\bm{k}}}{2}\\
-e^{\frac{i}{2}\gamma_{\bm{k}}}\text{cos}\frac{\varphi_{\bm{k}}}{2}
\end{pmatrix},\quad u_{2}(\bm{k})=\begin{pmatrix}e^{-\frac{i}{2}\gamma_{\bm{k}}}\text{cos}\frac{\varphi_{\bm{k}}}{2}\\
e^{\frac{i}{2}\gamma_{\bm{k}}}\text{sin}\frac{\varphi_{\bm{k}}}{2}
\end{pmatrix},\label{eq:u}
\end{gather}
where tan$\gamma_{\bm{k}}=T_{2}(\bm{k})/T_{1}(\bm{k})$ and cos$\varphi_{\bm{k}}=T_{3}(\bm{k})/|\bm{T}(\bm{k})|$.
The Chern number of a band can be calculated by integrating the Berry
curvature $\Omega_{i}(\bm{k}),\,i=1,2$ over the Brillouin zone (BZ)
$C_{i}=(1/2\pi)\int_{\text{B}Z}\Omega_{i}(\bm{k})dk_{x}dk_{y}$, with
$\Omega_{i}(\bm{k})=\left[\bm{\nabla}_{\bm{k}}\times\bm{A}_{i}(\bm{k})\right]_{z}$,
and $\bm{A}_{i}(\bm{k})=i\langle u_{i}(\bm{k})|\nabla_{\bm{k}}|u_{i}(\bm{k})\rangle$
is the Berry connection of the band. For the checkerboard lattice
model, the Chern number is found to be $C_{1,2}=\pm1$. With a partially
filled topological flat band and in the presence of an electron-electron
interaction, the system would become a FCI, as demonstrated in Ref.~\cite{Neupert}.

\subsection{Interactions}

While a flat Chern band provides a playground for electrons, it is
the electron-electron interaction that drives the system to a FCI
state. The interaction is usually assumed to have the form of the
density-density coupling, which in general can be written as $\hat{h}_{\text{int}}=\sum_{\bm{i},\bm{j},\tau_{1},\tau_{2}}V^{\tau_{1}\tau_{2}}(\bm{R}_{i}-\bm{R}_{j})\hat{n}_{\bm{i}\tau_{1}}\hat{n}_{\bm{j}\tau_{2}}$
, where $\hat{n}_{\bm{i}\tau}$ is the particle number operator at
the $\tau$ sub-lattice of the unit cell $\bm{i}$. Since only the
partially filled flat band is relevant to the FCI, one can project
the interaction to the band and obtain $\hat{h}_{\text{int}}^{p}=1/N\sum_{\bm{k}_{1},\bm{k}_{2},\bm{q}}M(\bm{k}_{1},\bm{k}_{2};\bm{q})\hat{\rho}_{\bm{k}_{1},\bm{q}}\hat{\rho}_{\bm{k}_{2},-\bm{q}}$,
where $\hat{\rho}_{\bm{k},\bm{q}}=\hat{d}_{\bm{k+q}}^{\dagger}\hat{d}_{\bm{k}}$,
$\hat{d}_{\bm{k}}$ ($\hat{d}_{\bm{k}}^{\dagger}$) is the annihilation
(creation) operator for a Bloch state in the topological flat band,
and $N$ is the total number of unit cells. The interaction matrix
element $M(\bm{k}_{1},\bm{k}_{2};\bm{q})$ can be written as:

\begin{align}
M(\bm{k}_{1},\bm{k}_{2};\bm{q}) & =\sum_{\tau_{1},\tau_{2}}V_{\bm{q}}^{\tau_{1}\tau_{2}}u_{1,\tau_{1}}^{\ast}(\bm{k}_{1}+\bm{q})\nonumber \\
 & \times u_{1,\tau_{1}}(\bm{k}_{1})u_{1,\tau_{2}}^{\ast}(\bm{k}_{2}-\bm{q})u_{1,\tau_{2}}(\bm{k}_{2})
\end{align}
where $u_{1,\tau}$ is the $\tau$-component of the eigenvector Eq.~(\ref{eq:u}),
$V_{\bm{q}}^{\tau_{1}\tau_{2}}=\sum_{\bm{R}}V^{\tau_{1}\tau_{2}}(\bm{R})\exp(-i\bm{q}\cdot\bm{R})$,
and we have assumed that the band $1$ is partially filled.

In literatures, the interaction is usually assumed to be of a short-range
one which only couples between NNs. It has the form: 

\begin{equation}
V^{\tau_{1}\tau_{2}}(\bm{R}_{i}-\bm{R}_{j})=\begin{cases}
U_{1} & \bm{i}\tau_{1},\,\bm{j}\tau_{2}\in NN\\
0 & \mathrm{others}
\end{cases},\label{eq:V1}
\end{equation}
In the checkerboard lattice model, it corresponds to:

\begin{equation}
V_{\bm{q}}=\begin{bmatrix}0 & U_{1}(\bm{q})\\
U_{1}^{\ast}(\bm{q}) & 0
\end{bmatrix},\label{eq:Vq1}
\end{equation}
where $U_{1}(\bm{q})=U_{1}(1+e^{-iq_{1}}+e^{-iq_{2}}+e^{-i(q_{1}+q_{2})/2})$.

While the short-range interaction is convenient for numerical simulations,
it is nevertheless very different from interactions in real systems.
In real materials, the interaction between two electrons that are
spatially far apart should be the Coulomb interaction: 

\begin{equation}
V^{\tau_{1}\tau_{2}}(\bm{R}_{i}-\bm{R}_{j})=\frac{U_{2}}{|\bm{r}_{\bm{i}}^{\tau_{1}}-\bm{r}_{\bm{j}}^{\tau_{2}}|},\label{eq:V2}
\end{equation}
where $\bm{r}_{\bm{i}}^{\tau_{1}}$ represents the real-space position
of the given lattice site. The interacting potential corresponds to:
\begin{equation}
V_{\bm{q}}=\begin{bmatrix}U_{2}(\bm{q}) & U_{2}^{\prime}(\bm{q})\\
U_{2}^{\prime\ast}(\bm{q}) & U_{2}(\bm{q})
\end{bmatrix},\label{eq:Vq2}
\end{equation}
where $U_{2}(\bm{q})=2\text{\ensuremath{\pi}}U_{2}\sum_{\bm{G}}|\bm{q}+\bm{G}|^{-1}$,
$U_{2}^{\prime}(\bm{q})=2\text{\ensuremath{\pi}}U_{2}\textrm{exp}(-i\bm{q}\cdot\bm{\tau})\sum_{\bm{G}}(-1)^{m+n}|\bm{q}+\bm{G}|^{-1}$,
$\bm{\tau}=(\frac{1}{2},\frac{1}{2})$ is a vector from A site to
B site, and the summation is over the reciprocal lattice vectors $\bm{G}=2\pi(m,n)$.

Moreover, one expects that the electron-electron interaction should
deviate from the Coulomb interaction at short distances. This is because
an electron in a lattice site of the tight-binding model is actually
corresponded to a finite size electron cloud with a spatial distribution
instead of an ideal point charge. To take account of the deviation,
we introduce an interaction which mixes the short-range interaction
and the Coulomb interaction. It reads,

\begin{equation}
V^{\tau_{1}\tau_{2}}(\bm{R}_{i}-\bm{R}_{j})=\begin{cases}
U_{1}, & \bm{i}\tau_{1},\bm{j}\tau_{2}\in NN\\
\frac{U_{2}}{|\bm{r}_{\bm{i}}^{\tau_{1}}-\bm{r}_{\bm{j}}^{\tau_{2}}|}, & \bm{i}\tau_{1},\bm{j}\tau_{2}\notin NN
\end{cases}.\label{eq:V3}
\end{equation}
The interaction is an interpolation between the short-range interaction
and the Coulomb interaction, and the ratio $U_{2}/U_{1}$ controls
the relative strengths of the two components. The interaction becomes
the pure Coulomb interaction when $U_{2}/U_{1}=1/\sqrt{2}$, while
the short-range component becomes more prominent when $U_{2}/U_{1}$
deviates from the ratio. 

The interactions introduced in Eq.~(\ref{eq:V1}), Eq. (\ref{eq:V2})
and Eq. (\ref{eq:V3}) are three representative forms which we will
survey in this paper. We will determine optimal gauges corresponding
to them and demonstrate how the form of interaction affects the construction
of the trial ground state wave function.

\subsection{Methods}

We determine the optimal gauge by using the variational principle
of interaction energy proposed by Zhang \textit{et al.}~\cite{Zhang2016}.
The gauge is represented by a function $\theta(\bm{k})$, which assigns
a $U(1)$ phase to each of the Bloch states in the BZ. The gauge determines
the spatial distributions of the 2D localized Wannier functions which
facilitate the mapping from Landau levels to a FCI lattice model,
and acts as variational parameters for the trial ground state wave
function. Accordingly, $\theta(\bm{k})$ are determined by the variational
principle of ground state energy, which is equivalent to minimizing
the interaction energy functional~\cite{Zhang2016}:
\begin{multline}
E_{\text{int}}[\theta(\bm{k})]=\frac{1}{N}\sum_{\bm{k}_{1},\bm{k}_{2},\bm{q}}M(\bm{k}_{1},\bm{k}_{2};\bm{q})\Pi(\bm{k}_{1}-\bm{k}_{2};\bm{q})\\
\times e^{-i\left[\theta(\bm{k}_{1}+\bm{q})-\theta(\bm{k}_{1})+\theta(\bm{k}_{2}-\bm{q})-\theta(\bm{k}_{2})\right]}\label{eq: energy functional}
\end{multline}
where $\Pi(\bm{k}_{1}-\bm{k}_{2};\bm{q})$ is the two-particle correlation
function of the FQH state to be mapped, and can be determined by~\cite{Zhang2016}:
\begin{align}
\Pi(\bm{k}_{1}-\bm{k}_{2};\bm{q}) & =\nu^{2}(\delta_{\bm{q},0}-\delta_{\bm{k}_{1}-\bm{k}_{2}+\bm{q},0})\nonumber \\
 & +e^{-i(\bm{A}_{\bm{k}_{1}+\bm{q}}^{L}-\bm{A}_{\bm{k}_{2}}^{L})\cdotp\bm{q}}\Pi'(\bm{k}_{1}-\bm{k}_{2};\bm{q}),
\end{align}

\begin{align}
\Pi'(\bm{k};\bm{q}) & =\frac{1}{N}\sum_{\bm{R}}\Pi'(\bm{R};\bm{q})e^{-i\bm{k}\cdotp\bm{R}},\\
\Pi'(\bm{R};\bm{q}) & =2\nu^{2}\exp\left[-i\bm{q}\cdotp\bm{R}-\frac{1}{2l_{M}^{2}}(\bm{R}-\hat{z}\times\bm{q}C_{1}l_{M}^{2})^{2}\right]\nonumber \\
 & \times\sum_{k=0}^{\infty}c_{2k+1}L_{2k+1}\left(\frac{1}{l_{M}^{2}}(\bm{R}-\hat{z}\times\bm{q}C_{1}l_{M}^{2})^{2}\right)
\end{align}
where $\nu$ is the filling factor, $\bm{R}$ denotes a lattice vector,
$\bm{A}_{\bm{k}}^{L}=(-C_{1}k_{2}/2\pi,0)$, $l_{M}=1/\sqrt{2\pi}$,
$L_{n}(x)$ is the Laguerre function, and the coefficients $c_{2k+1}$
for $\nu=1/3$ and $1/5$ can be found in Ref.~\cite{GMP}. 

We define the correlation energy $E_{\text{corr}}=E_{\text{int}}-E_{\text{HF}}$,
where $E_{\text{HF}}$ is the mean-field interacting energy determined
by the Hartree-Fock approximation, and can be written as:
\begin{equation}
E_{\text{HF}}=\frac{1}{N}\sum_{\bm{k}_{1},\bm{k}_{2}}M(\bm{k}_{1},\bm{k}_{2};0)\nu-M(\bm{k}_{1},\bm{k}_{2};\bm{k}_{2}-\bm{k}_{1})\nu(1-\nu),
\end{equation}
which is independent of the choice of the gauge. Different choices
of the gauge affect how electrons are correlated locally, and give
rise to different correlation energies. We adopt the correlation energy
as an indicator for the quality of a trial ground state wave function.

The function $\theta(\bm{k})$ can be used to construct the projected
Wannier functions~\cite{Zhang2016}:
\begin{equation}
w_{\tau}(\bm{r},\bm{R})=\frac{1}{\sqrt{N}}\sum_{\bm{k}}\varphi_{1\bm{k}}(\bm{r})u_{1,\tau}^{\ast}(\bm{k}){\text{exp}{(-i\bm{k}\cdot\bm{R}-i\theta(\bm{k}))}}
\end{equation}
where $\varphi_{1\bm{k}}(\bm{r})$ is a magnetic Bloch wave function
of the LLL with the same Chern number $C_{1}$ as the partially filled
flat Chern band. The gauge of $\varphi_{1\bm{k}}(\bm{r})$ and $u_{1}(\bm{k})$
should be regularized to satisfy the same quasi-periodic conditions
for $\psi_{\bm{k}}\equiv\varphi_{1\bm{k}}\,\mathrm{or}\,u_{1}(\bm{k})$:
\begin{align}
\psi_{\bm{k}+\bm{K}_{1}} & =\psi_{\bm{k}},\nonumber \\
\psi_{\bm{k}+\bm{K}_{2}} & =\psi_{\bm{k}}\exp\left(iC_{1}k_{1}\right),\label{eq:bc}
\end{align}
where $\bm{K}_{1}=2\pi(1,0)$, $\bm{K}_{2}=2\pi(0,1)$. The Wannier
functions are spatially localized and can be employed to map a Landau
level to the partially filled flat Chern band.

The mapping facilitated by the Wannier functions is equivalent to
mapping the magnetic Bloch wave functions $\varphi_{1\bm{k}}$ to
the lattice Bloch wave functions $u_{1}(\bm{k})e^{i\theta(\bm{k})}$,
through which $\{\theta(\bm{k})\}$ become variational parameters
of the trial ground state wave function. While mappings with different
$\{\theta(\bm{k})\}$ lead to the same kinetic part of a FCI hamiltonian,
their interaction energies will be different due to different density
distributions of the Wannier functions. A direct application of the
variational principle of ground state energy immediately leads to
the variational principle dictated by the interaction energy functional
Eq.~(\ref{eq: energy functional}). 

\subsection{Numerical implementation}

We implement our numerical calculation in a discretize BZ with a $21\times21$
mesh of $\bm{k}$ points. In the real space, it corresponds to a finite
size lattice with $21\times21$ unit cells and periodic boundary conditions.
The size is much larger than the coherence length beyond which $\Pi^{\prime}(\bm{R},\bm{q})$
approaches a constant (See Fig.~3 of Ref.~~\cite{Zhang2016}).
We focus on the case of a $\nu=1/3$ filled band which is a counterpart
of the FQH state with the filling factor $\nu=1/3$.

The regularized Bloch wave functions consistent with the quasi-periodic
conditions Eq.~(\ref{eq:bc}) can be explicitly chosen. For the magnetic
Bloch wave functions, we adopt the form:
\begin{align}
\varphi_{1\bm{k}}(\bm{r}) & =(\frac{\sqrt{2}}{N})^{\frac{1}{2}}\sum_{m\in\mathbb{Z}}\text{exp}\left[-iC_{1}k_{1}m+i(k_{2}+2\pi m)y\right]\nonumber \\
 & \times\text{exp}\left[-\pi\left(x+k_{2}/2\pi+m\right)^{2}\right].
\end{align}
The lattice Bloch wave function $u_{1}(\bm{k})$ can be regularized
in the discretize BZ. Starting from $u_{1}(\bm{k}=0)$, the phases
for the wave functions along the $k_{2}$ axis are chosen to satisfy
the condition:
\begin{align}
\textrm{Im}[\langle u_{0,n}|u_{0,n+1}\rangle] & =0,\,\textrm{Re}[\langle u_{0,n}|u_{0,n+1}\rangle]>0,
\end{align}
for $n=0\dots M-2$, where $u_{m,n}$ denotes the lattice Bloch wave
function $u_{1}(\bm{k})$ at the mesh point $(m,n)$ of the discretized
BZ, and $M=21$ is the size of the mesh. Then, the phases of other
Bloch wave functions are chosen to satisfy the condition: 
\begin{multline}
\textrm{Im}[\langle u_{m,n}|u_{m+1,n}\rangle]=0,\textrm{Re}[\langle u_{m,n}|u_{m+1,n}\rangle]>0,
\end{multline}
for $m=0\dots M-2,\,n=0\dots M-1$. Finally, we make a global gauge
transformation: 
\begin{equation}
u_{m,n}\rightarrow u_{m,n}\textrm{exp}\left[-\frac{i}{M}\left(n\delta_{0}+m\delta_{n}\right)\right],
\end{equation}
where $\delta_{0}=\arg\langle u_{0,0}|u_{0,M-1}\rangle$, and $\delta_{n}=\langle u_{0,n}|u_{M-1,n}\rangle$.
These regularized wave functions define our initial gauge, which corresponds
to $\theta(\bm{k})=0$ in the energy functional Eq.~(\ref{eq: energy functional}).
It turns out that the initial gauge is exactly the gauge adopted in
Qi's approach~\cite{Qi2011,Zhang2016}.

To determine $\theta(\bm{k})$ that minimizes the energy functional,
we employ the steepest descent algorithm. We run multiple iterations
starting from different initial values, and check whether they converge
to the same result. In this way, convergences to global minimums are
guaranteed.

\section{Results}

\subsection{Optimal gauge for the short-range interaction}

\begin{figure}
\includegraphics[width=1\columnwidth]{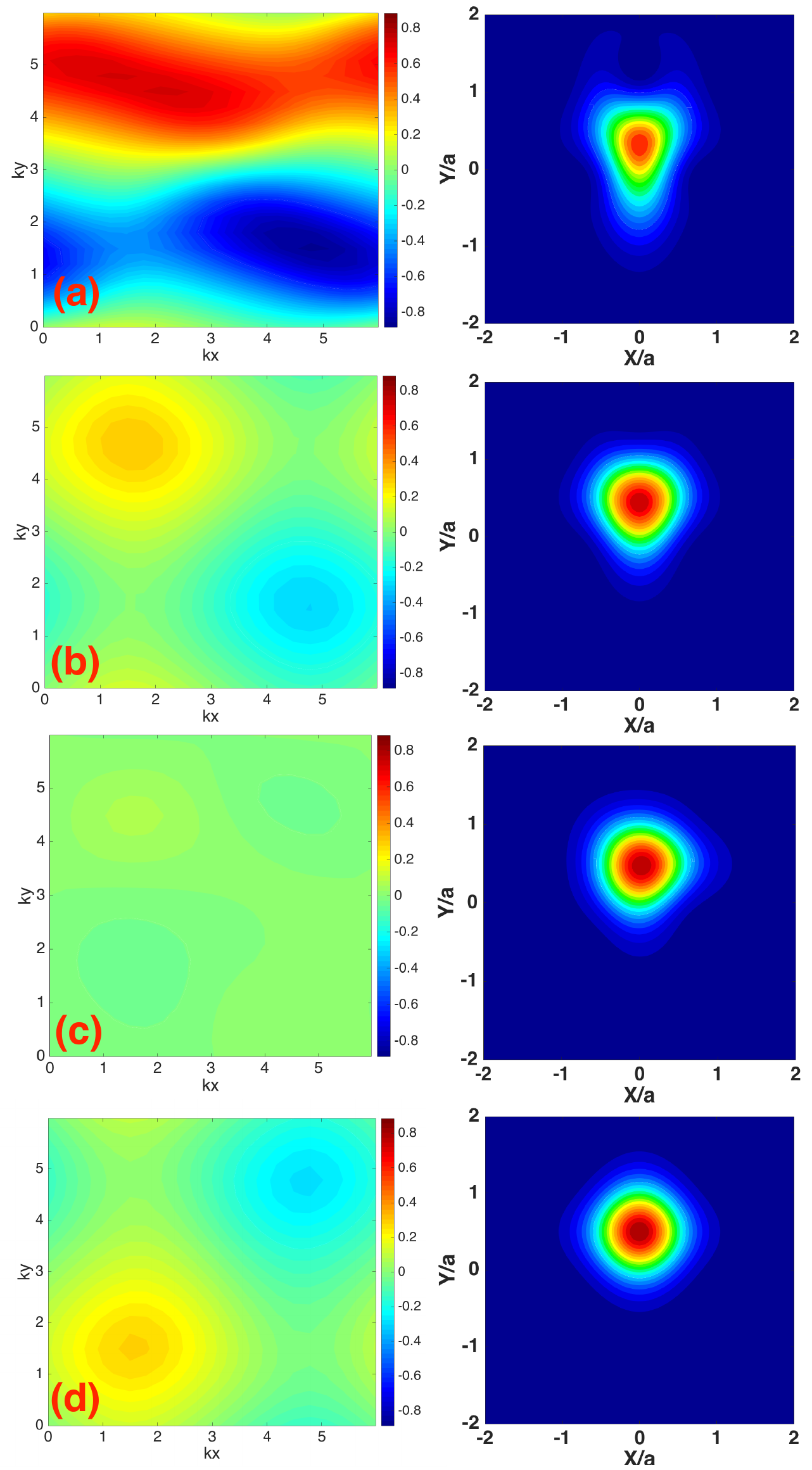} \caption{(Color online) Distributions of $\theta(\bm{k})$ (left) in the BZ
and the spatial distributions of the corresponding projected Wannier
functions at A-site (right) for different gauges. (a) The optimal
gauge for the short-range interaction defined in Eq.~(\ref{eq:V1});
(b) Wu\textit{ et al.}'s gauge; (c) The maximally localized gauge;
(d) The symmetric gauge. Qi's gauge corresponds to $\theta(\bm{k})=0$,
and is very close to the maximally localized gauge.}

\label{fig: gauge_sr}
\end{figure}

We determine the optimal gauge for the short-range interaction defined
in Eq.~(\ref{eq:V1}). The distribution of $\theta(\bm{k})$ in the
BZ and the spatial distribution of corresponding projected Wannier
functions at the A-site are shown in the left panel and right panel
of Fig.~2(a), respectively. For comparison, we also show results
for Wu\textit{ et al.}'s gauge~\cite{Wu2012}, the maximally localized
gauge that gives rise to a projected Wannier function maximally localized
in the real space, as well as a symmetric gauge which give rises to
a $C_{4}$ symmetric Wannier function. The last two gauges are defined
in Ref.~\cite{Zhang2016}. We find that $\theta(\bm{k})$ of the
optimal gauge has an amplitude $\sim\pi/4$. It shows a significant
deviation from the original gauge proposed by Qi, \emph{i.e.} $\theta(\bm{k})=0$.
In comparison, $\theta(\bm{k})$ of Wu \textit{et al}.'s gauge has
a smaller amplitude $\sim\pi/10$, with a distribution in the BZ qualitatively
similar to that of the optimal gauge, \emph{i.e.}, both of them have
a peak and a valley located in the same regions of the BZ. It indicates
that Wu\textit{ et al.}'s gauge, while not optimal, is nevertheless
better than Qi's gauge. On the other hand, the maximally localized
gauge yields $\theta(\bm{k})$ with an amplitude $10^{-2}$. It is
not identical but very close to Qi's gauge. Reference~\cite{Zhang2016}
proves that the maximally localized gauge is the optimal gauge for
a soft and isotropic interaction. Hence, Qi's gauge should be good
for the case, but is not a good choice for the short-range interaction.
Finally, the symmetric gauge has a distribution of $\theta(\bm{k})$
distinct from the optimal gauge.

We also show the spatial distributions of the corresponding Wannier
functions at the A-site for different gauges in the right panel of
Fig.~\ref{fig: gauge_sr}. The Wannier function of the B-site can
be obtained by  a reflection with respect to the diagonal $y=x$.
We observe that both the Wannier functions for the optimal gauge and
Wu\textit{ et al.}'s gauge have the mirror symmetry with respect to
the $y$-axis, while the one for Qi's gauge (or the maximally localize
gauge) has the mirror symmetry with respect to the diagonal $y=-x$.
It indicates that Qi's gauge is qualitatively different from the optimal
one. Comparing Wu \emph{et al.}'s gauge and the optimal gauge, we
find that the former is more localized spatially, and the latter is
elongated along the $y$-direction.

\begin{table}
\caption{Comparison of the interaction energy and the correlation energy per
electron for different gauges. Both results for the short-range interaction
and the Coulomb interaction are shown. The unit of the energy is $U_{1}$
($U_{2})$ for the short-range (Coulomb) interaction.}
\centering %
\begin{tabular}{|c|c|c|c|c|c|c|}
\hline 
 & \multicolumn{3}{c|}{Short-range interaction} & \multicolumn{3}{c|}{Coulomb interaction}\tabularnewline
\hline 
gauge  & $E_{\text{int}}$  & $E_{\text{corr}}$  & $\frac{\Delta E_{\text{corr}}}{E_{\text{corr}}^{OP}}$ & $E_{\text{int}}$  & $E_{\text{corr}}$  & $\frac{\Delta E_{\text{corr}}}{E_{\text{corr}}^{OP}}$\tabularnewline
\hline 
Qi  & 0.2893  & 0.0697  & 17.74\% & 11.7061 & 0.1660  & 2.72\%\tabularnewline
Wu \textit{et al.}  & 0.2851  & 0.0655  & 10.64\% & 11.7032 & 0.1631 & 0.93\%\tabularnewline
symmetric & 0.2921 & 0.0725 & 22.47\% & 11.7079 & 0.1678 & 3.84\%\tabularnewline
Optimal & 0.2788  & 0.0592  & /  & 11.7018 & 0.1616  & / \tabularnewline
\hline 
\end{tabular}

\label{Table: energy}
\end{table}

Table \ref{Table: energy} shows the comparison of the interaction
energies and correlation energies for different gauges in the short-range
interaction. Compared to the optimal gauge, the other three gauges
show significantly higher ($>10\%$) correlation energies. Among them,
Wu \textit{et al.}\textit{\emph{'s gauge is closest to the optimal
one. The symmetric gauge turns out to be the worst since it is based
on an }}\textit{ad hoc}\textit{\emph{ requirement that the Wannier
function should have the same point group symmetry as its hosting
lattice.}}

\subsection{The optimal gauge for the Coulomb interaction}

\begin{figure}
\centering \includegraphics[width=1\columnwidth]{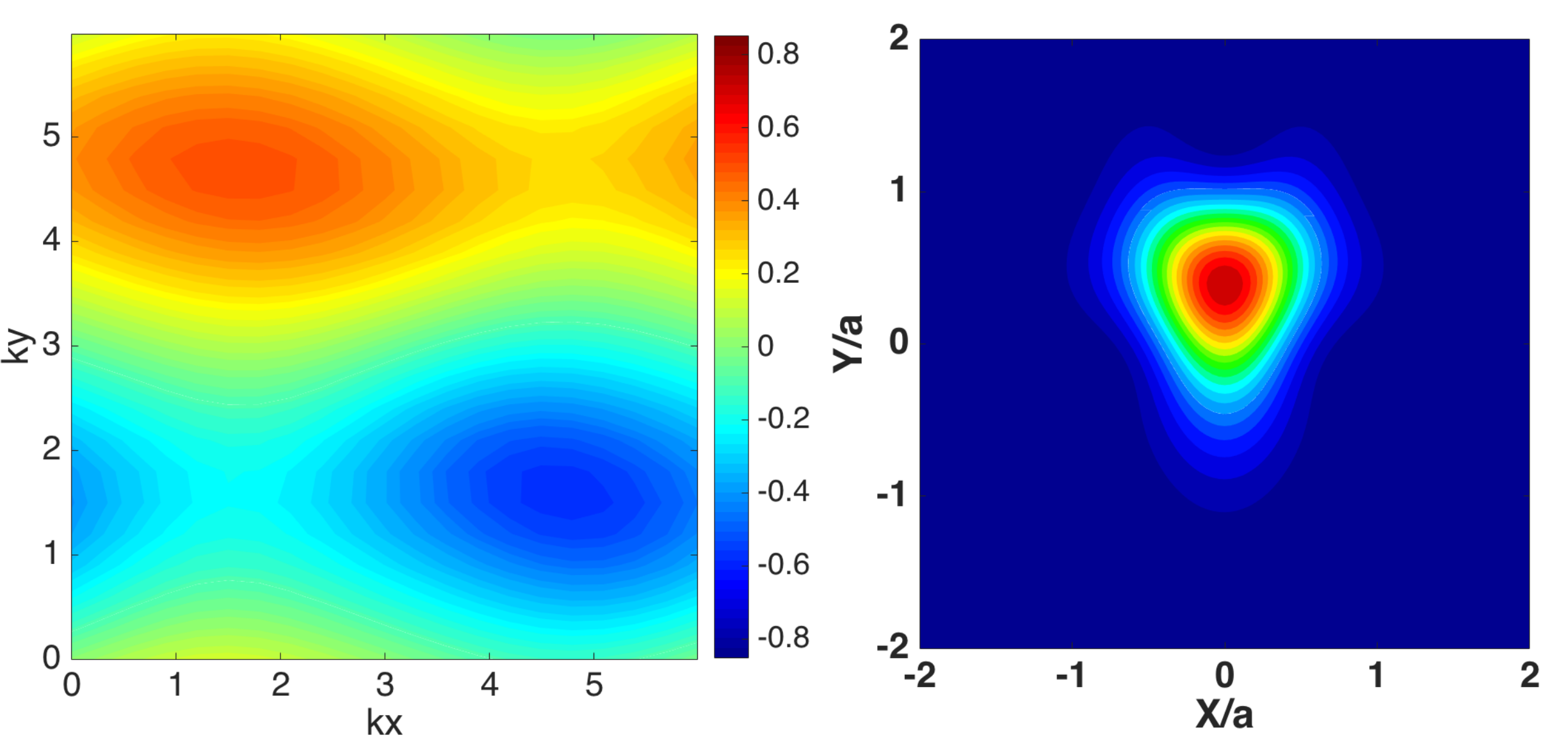} \caption{(Color online) Distribution of $\theta(\bm{k})$ in the BZ (left)
and the spatial distribution of the Wannier function at the A-site
(right) of the optimal gauge for the Coulomb interaction. }

\label{fig: gauge_lr}
\end{figure}

We also determine the optimal gauge for the Coulomb interaction, shown
in Fig.~\ref{fig: gauge_lr}. Compared to the optimal gauge for the
short-range interaction, the Wannier function becomes more localized
spatially and shorten along the $y$ direction. The observation is
a clear indication that the optimal gauge depends on the form of interaction. 

In Table \ref{Table: energy}, we also show that the interaction energy
and correlation energy per electron for the Coulomb interaction. We
observe that the difference in the energy between different gauges
becomes much smaller. This is not surprising because different gauges
only change the density distribution of the Wannier function. The
local change can only affects the short-range correlations of electrons,
and has a relatively minor effect when the interaction is of a long-range
one.

\subsection{Evolution of the optimal gauge with interactions }

\begin{figure}
\centering \includegraphics[width=1\columnwidth]{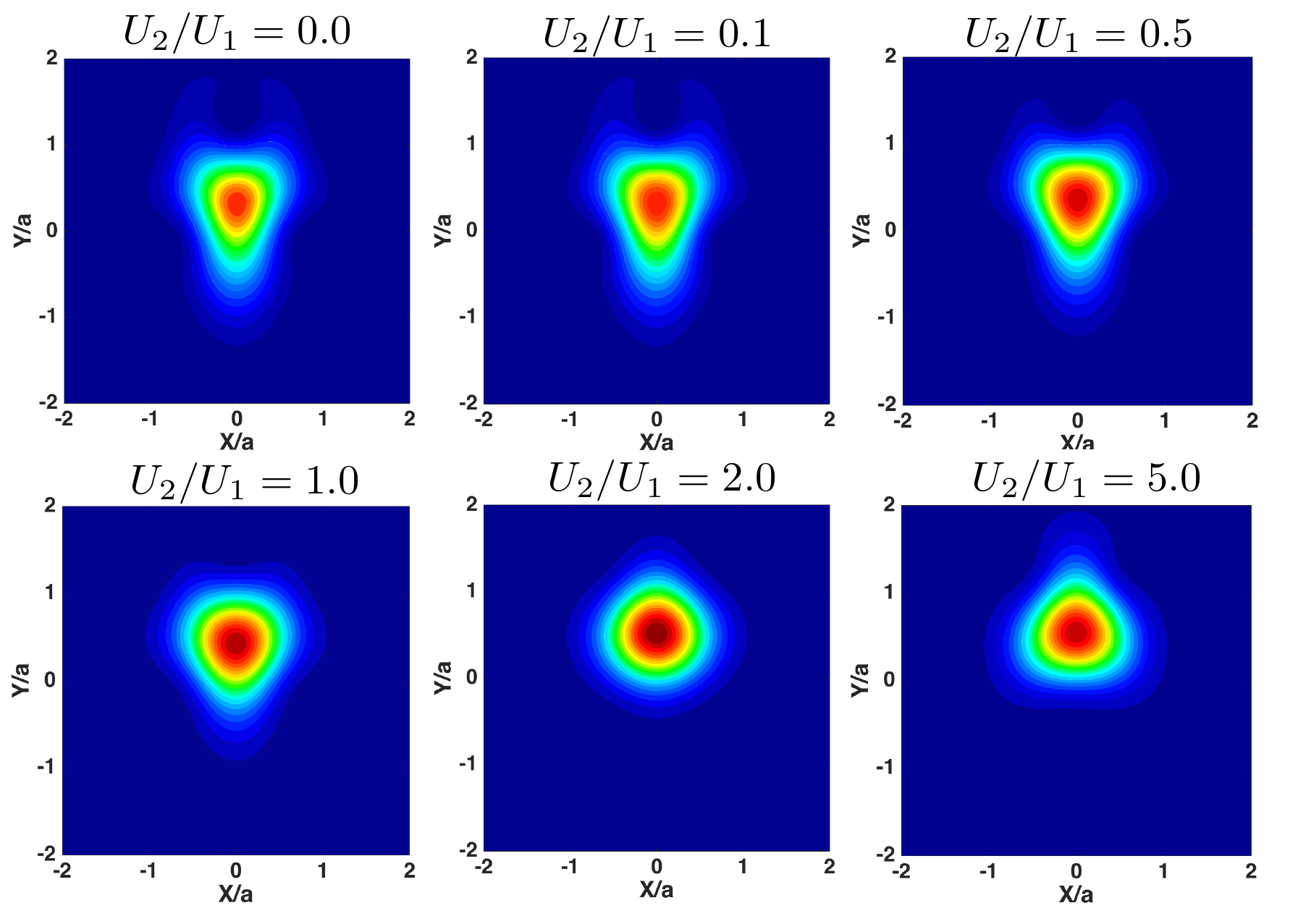} \caption{(Color online) Evolution of the Wannier function of the optimal gauge
for the mixed form of interaction Eq.~(\ref{eq:V3}) with the ratio
$U_{2}/U_{1}$ varied from 0.0 to 5.0. $U_{2}/U_{1}=0$ corresponds
to the short-range interaction Eq.~(\ref{eq:V1}), and $U_{2}/U_{1}=1.0$
corresponds to the case where the Wannier function is very close to
that in Wu \textit{et al}.'s gauge.}

\label{fig: evolution}
\end{figure}

After establishing the dependence of the optimal gauge on the form
of interaction, we proceed to see how the optimal gauge evolves with
different forms of interactions. We adopt the mixed form of the interaction
Eq.~(\ref{eq:V3}), and determine optimal gauges for different values
of $U_{2}/U_{1}$. Figure~\ref{fig: evolution} shows the evolution
of corresponding Wannier functions. We observe that the overall shapes
of the Wannier functions undergo clearly visible changes, from a elongated
triangle pointing downward for the short-range interaction ($U_{2}/U_{1}=0$),
to a nearly equilateral triangle pointing downward when $U_{2}/U_{1}=1$,
and to triangles pointing upward when $U_{2}/U_{1}$ is further increased.
Actually, the Wannier function for the interaction ($U_{2}/U_{1}\sim1)$
is very close to that in Wu \textit{et al}.'s gauge shown in Fig.~\ref{fig: gauge_sr}(b).

\begin{figure}
\includegraphics[width=1\columnwidth]{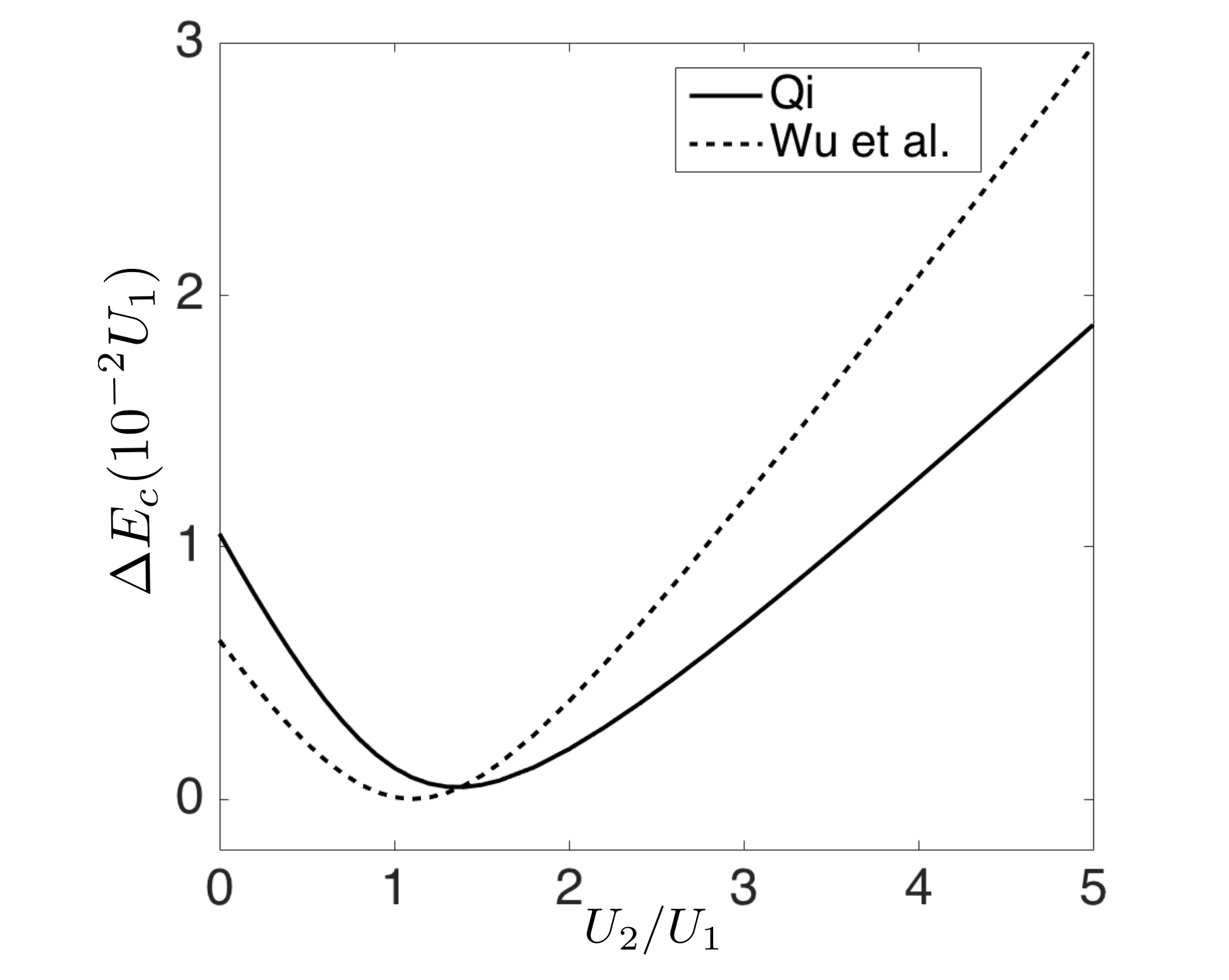}\caption{Correlation energies versus $U_{2}/U_{1}$ of Qi's gauge and Wu\textit{
et al.}'s gauge relative to the optimal one. }

\label{fig: energy}
\end{figure}

In Fig.~\ref{fig: energy}, we show the correlation energy corresponding
to Qi's gauge and Wu\textit{ et al.}'s gauge relative to that of the
optimal gauge versus the ratio $U_{2}/U_{1}$. We see that the correlation
energies of both the gauges can be close to that of the optimal gauge
when the strengths of the short-range interaction and the Coulomb
interaction are approximately equal ($U_{2}/U_{1}\sim1$), but deviate
significantly when the short-range component of the interaction becomes
more prominent ($U_{2}/U_{1}\ll1$ or $\gg1$). This is not surprising
because our variational degrees of freedom $\theta(\bm{k})$ affects
mainly short-range correlations in the trial ground state wave function
by modifying the spatial distribution of the Wannier functions. Its
effect to a system with a long-range interaction like the Coulomb
interaction would be minimal. We also observe that while Wu\textit{
et al.}'s gauge can be very close to the optimal gauge when $U_{2}/U_{1}\approx1$,
it actually becomes worse than Qi's gauge when $U_{2}/U_{1}\gg1$
. This is because in the regime, both the gauges are qualitatively
different from the optimal gauge, and Qi's gauge is actually relatively
closer to the optimal one, as evident from Fig.~\ref{fig: evolution}
and Fig.~\ref{fig: gauge_sr}(b, c).

\section{Concluding remark}

In conclusion, we have demonstrated the evolution of the optimal gauge
for constructing the trial wave function of the FCI in the checkerboard
model. It clearly indicates that the optimal gauge is not only determined
by the kinetic property of a flat Chern band, but also the form of
interaction. We also compare the optimal gauge with those proposed
by Qi and Wu \textit{et al}.. We find that both the gauges deviate
from the optimal one when the short-range component in the interaction
becomes more prominent, although Wu \textit{et al}.'s gauge can be
very close to the optimal gauge for a certain mixture of the short-range
interaction and the Coulomb interaction ($U_{2}/U_{1}\sim1$).


\begin{thebibliography}{10}
\bibitem{Tsui} D. C. Tsui, H. L. Stormer and A. C. Gossard, Phys.
Rev. Lett. \textbf{48}, 1559 (1982).

\bibitem{Klitzing1980} K. v. Klitzing, G. Dorda and M. Pepper, Phys.
Rev. Lett. \textbf{45}, 494 (1980).

\bibitem{TKNN} D. J. Thouless, M. Kohmoto, M. P. Nightingale and
M. den Nijs, \textbf{49}, 405 (1982).

\bibitem{Laughlin1983} R. B. Laughlin, Phys. Rev. Lett. \textbf{50},
1395 (1983).

\bibitem{Jain1989} J. K. Jain, Phys. Rev. Lett. \textbf{63}, 199
(1989).

\bibitem{JainCF} J. K. Jain, \textit{Composite Fermions }(Cambridge
University Press, Cambridge, England, 2007).

\bibitem{Haldane1983} F. D. M. Haldane, Phys. Rev. Lett. \textbf{51},
605 (1983).

\bibitem{MR} G. Moore and N. Read, Nucl. Phys. B \textbf{360}, 362(1991)
See also N. Read and G. Moore, Prog. Theor. Phys. Suppl. \textbf{107},
157 (1992).

\bibitem{NayakRMP} C. Nayak, S. H. Simon, A. Stern, M. Freedman and
S. Das Sarma, Rev. Mod. Phys. \textbf{80}, 1083 (2008).

\bibitem{Sheng} D. N. Sheng, Z. Gu, K. Sun, and L. Sheng, Nature
Commun. \textbf{2}, 389 (2011).

\bibitem{SunKai} K. Sun, Z. Gu, H. Katsura, and S. Das Sarma, Phys.
Rev. Lett. \textbf{106}, 236803 (2011).

\bibitem{Neupert} T. Neupert, L. Santos, C. Chamon, and C. Mudry,
Phys. Rev. Lett. \textbf{106}, 236804 (2011).

\bibitem{Tang} E. Tang, J.-W. Mei, and X.-G. Wen, Phys. Rev. Lett.
\textbf{106}, 236802 (2011).

\bibitem{Wangyifei} Y.-F.Wang, Z.-C. Gu, C.-D. Gong, and D. N. Sheng,
Phys. Rev. Lett. \textbf{107}, 146803 (2011).

\bibitem{Bernevigprx} N. Regnault and B. A. Bernevig, Phys. Rev.
X \textbf{1}, 021014 (2011).

\bibitem{Parameswaran} S. A. Parameswaran, R. Roy and S. L. Sondhi,
C. R. Physique \textbf{14}, 816 (2013).

\bibitem{LiuZhaoreview} E. J. Bergholtz and Z. Liu, Int. J. Mod.
Phys. B \textbf{27}, 1330017 (2013).

\bibitem{NeupertFTI} T. Neupert, C. Chamon, T. Iadecola, L. H. Santos
and C. Mudry4, Phys. Scr. T164 (2015) 014005 (9pp).

\bibitem{Qi2011} X. L. Qi, Phys. Rev. Lett, \textbf{107},126803 (2011).

\bibitem{Wu2012} Y. L. Wu, N. Regnault and B. A. Bernevig, Phys.
Rev. B \textbf{86}, 085129 (2012).

\bibitem{Wu2013} Y. L. Wu, N. Regnault and B. A. Bernevig, Phys.
Rev. Lett. \textbf{110}, 106802 (2013).

\bibitem{Wu2014} Y. L. Wu, N. Regnault and B. A. Bernevig, Phys.
Rev. B \textbf{89}, 155113 (2014).

\bibitem{Zhang2016} Y. H. Zhang and J. R. Shi, Phys. Rev. B \textbf{93},
165129 (2016).

\bibitem{LiuzhaoHighC} Z. Liu, E. J. Bergholtz, H. Fan, and A. M.
Lauchli, Phys. Rev. Lett. \textbf{109}, 186805 (2012).

\bibitem{Sterdyniak} A. Sterdyniak, C. Repellin, B. Andrei Bernevig,
and N. Regnault, Phys. Rev. B \textbf{87}, 205137 (2013).

\bibitem{Trescher} M. Trescher and E. J. Bergholtz, Phys. Rev. B
\textbf{86}, 241111 (2012).

\bibitem{Wangfa} F. Wang and Y. Ran, Phys. Rev. B \textbf{84}, 241103
(R) (2011).

\bibitem{WangyifeiHighC} Y.-F. Wang, H. Yao, C.-D. Gong, and D. N.
Sheng, Phys. Rev. B \textbf{86}, 201101 (2012).

\bibitem{Yangshuo} S. Yang, Z. C. Gu, K. Sun and S. Das Sarma, Phys.
Rev. B \textbf{86}, 241112 (2012).

\bibitem{Cooper} N. R. Cooper and R. Moessner, Phys. Rev. Lett. \textbf{109},
215302 (2012).

\bibitem{GMP} S. M. Girvin, A. H. MacDonald, P. M. Platzman, Phys.
Rev. B \textbf{33}, 2481 (1986).
\end{thebibliography}
\end{document}